\documentstyle[aps,prl,multicol,epsf,amssymb]{revtex}

\def\be{\begin{equation}}
\def\en{\end{equation}}

\begin{document}
\draft
\bibliographystyle{prsty}
\title{Simulating Particle Dispersions in Nematic Liquid-Crystal Solvents}
\author{Ryoichi Yamamoto\cite{byline}}
\address{Department of Chemistry, University of Cambridge, 
Lensfield Road, Cambridge CB2 1EW, UK}
\date{\today}
\maketitle

\begin{abstract}

A new method is presented for mesoscopic simulations of particle 
dispersions in nematic liquid crystal solvents.
It allows efficient first-principle simulations of the dispersions 
involving many particles with many-body interactions mediated by
the solvents.
A simple demonstration is shown for the aggregation process of 
a two dimensional dispersion.

\end{abstract}


\begin{multicols}{2}
\narrowtext


Dispersions of small particles in host fluids such as 
colloidal suspensions and emulsions are of considerable 
technological importance, 
and often appear in our everyday life in
paints, foods, and drugs.
Many kinds of exotic interactions between particles 
mediated by the host fluids are possible, including 
screened Coulombic \cite{SC}, depletion \cite{SC}, 
fluctuation induced \cite{casimir}, 
and surface induced \cite{borstnik} forces.
A striking example occurs when spherical particles are
immersed in a liquid-crystal solvent in the nematic phase.
For a single particle, the orientation of the solvent molecules 
is distorted due to the anchoring of the solvent molecules at the
particle surface.
Extensive studies have been done on this effect, 
and several characteristic configurations of the nematic 
field around a spherical particle have been identified 
\cite{terentjev-sat,ramaswamy,terentjev-sim,stark,mondain,gu}.
When the strength of anchoring is increased so that 
normal anchoring is preferred, the solvent changes from 
quadrupolar to dipolar symmetries around the particle.
When more than two particles are immersed in the solvent, 
long-range anisotropic interactions are induced between particles 
due to elastic deformations of the nematic field
\cite{poulin,terentjev-the,lubensky,lev}.
The anisotropic interactions can have a pronounced effect not only 
on the local correlations of the particles \cite{poulin}, 
but also on their phase behavior 
\cite{raghunathan,terentjev-exp,loudet,jyamamoto}
and on their mechanical properties \cite{terentjev-exp}.

Since analytical approaches for investigating these kinds of 
complex materials are extremely difficult, computer simulations 
are essential to investigate their static and dynamical properties.
In most dispersions, the host fluid molecules are much smaller and 
move much faster than the dispersed particles.
This enables us to assume that the host fluid is in local equilibrium 
for any given particle configurations, and thus it is usually a good
idea to use some coarse grained mesoscopic descriptions for the host
fluids rather than treating them as fully microscopic molecules \cite{MD}.
In the case of charged colloidal suspensions, 
a mesoscopic method for the first principle simulations can be derived 
by treating the counter ions as a {\it charge density} \cite{lowen}. 
For the particle dispersions in nematic solvents considered here, 
the mesoscopic coarse grained free energy for the nematic solvent 
is well known to be the Frank free energy \cite{frank},
and the total free energy ${\cal F}$ of the system 
consists of the following two parts; 
the bulk term ${\cal F}_{el}$ which presents 
elastic energy of the nematic and the surface term ${\cal F}_{s}$ which
determines anchoring of the nematic field at the particle surface.
Let {\bf n}({\bf r}) be the {\it director}, a common direction 
on which solvent molecules are aligned on average with a constraint 
$|{\bf n}({\bf r})|=1$.
${\cal F}$ can be given by functionals of $[{\bf n}({\bf r})]$ 
for a given particle 
configuration $\{{\bf R}_1\cdots{\bf R}_N\}$;
\begin{eqnarray}
&&{\cal F}([{\bf n}({\bf r})];\{{\bf R}_1\cdots{\bf R}_N\})\nonumber\\
&&=\frac{K}{2}\int d{\bf r}\left[(\nabla\cdot{\bf
n})^2+(\nabla\times{\bf n})^2\right]
+\frac{W}{2}\oint dS [1-({\bf n}\cdot\mbox{\boldmath$\nu$})^2],
\label{Frank1}
\end{eqnarray}
where $K$ is the Frank constant with the single
elastic constant approximation,
$W$ is the surface anchoring constant, 
and \mbox{\boldmath$\nu$} is the unit
vector normal to the colloidal surface \cite{terentjev-sim,stark}.
The saddle-splay elastic term \cite{stark} is not considered 
in Eq.(\ref{Frank1}).
The integral in the first term, ${\cal F}_{el}$, runs over the whole 
solvent volume excluding the particles, 
and that in the second term, ${\cal F}_{s}$, runs over all 
solvent-particle interfaces.
A simple scaling argument tells us ${\cal F}_{el}\propto Ka^{d-2}$
and ${\cal F}_{s}\propto Wa^{d-1}$ with $a$ and $d$ being 
the particle radius and the system dimension, respectively, thus
the physics should be determined by the ratio 
${\cal F}_{s}/{\cal F}_{el}\propto Wa/K$.
Although this type of free energy functional is sufficient for 
performing Monte Carlo simulations where only values of ${\cal F}$
are needed for a given particle configurations, it is not useful 
for molecular dynamics (MD) or Brownian type simulations 
because the coupling between solvent 
and the particles is given implicitly by limiting the integration space 
in both ${\cal F}_{el}$ and ${\cal F}_{s}$.
This produces mathematical singularities at the interface when one 
calculates the force, 
${\bf f}_i^{PS}=-\partial {\cal F}/\partial {\bf R}_i$, acting on each 
particle mediated by the nematic solvents.
Calculating the force is crucial for performing efficient
simulations of many particle systems.
Another serious problem of this type of functional 
is that in order to give correct boundary conditions at the 
particle-solvent interface, one has to use appropriate coordinates 
for performing grid-based numerical simulations rather than the 
usual Cartesian coordinates.
This is generally difficult for particles with non-spherical shapes or 
for systems involving many particles even when each particle has 
a spherical shape.
Also this makes the use of the periodic boundary condition difficult.

To overcome these problems, we have modified Eq.(\ref{Frank1}) by
using a smooth interface between the solvent and the particles 
so that the coupling is given explicitly in the integrants 
through the interface.
The new free energy functional we propose is 
\begin{eqnarray}
{\cal F}&([&{\bf q}({\bf r})];\{{\bf R}_1\cdots{\bf R}_N\})\nonumber\\
&=&\frac{K}{4R_c^2}\int d{\bf r}\left(1-\sum_{i=1}^N\phi_i({\bf
r})\right)\tanh\left[R_c^2(\nabla_\alpha q_{\beta \gamma})^2\right]\nonumber\\
&+&\frac{W\xi}{2}\int d{\bf r}\sum_{i=1}^N\left[\frac{d-1}{d}(\nabla_\alpha\phi_i)^2-(\nabla_\alpha\phi_i)(\nabla_\beta\phi_i) q_{\alpha\beta}\right] ,
\label{Frank2}
\end{eqnarray}
where $\alpha,\beta,\gamma\in x,y,z$ and the summation convention is used.
The explicit form of the interfacial profile $\phi_i$ between dispersed
particles and solvents is given by
\be
\phi_i({\bf r})
=\frac{1}{2}\left(\tanh\frac{a-|{\bf r}-{\bf
R}_i|}{\xi}+1\right),
\label{profile}
\en
with the particle radius $a$ and the interface thickness $\xi$.
Note that this reduces to Eq.(\ref{Frank1}) if 
$R_c,\xi\rightarrow0$.
Very recently, a similar idea of using smooth interface was proposed 
for treating the hydrodynamic forces acting on particles dispersed 
in simple liquids \cite{tanaka}.
In our case, the free energy is given by functionals of a traceless 
and symmetric second-rank tensor 
$
q_{\alpha\beta}({\bf r})=n_\alpha({\bf r}) n_\beta({\bf r})
-\delta_{\alpha\beta}/d 
$
rather than the director ${\bf n}({\bf r})$ to take into account the 
symmetry of the nematic director $+{\bf n}\leftrightarrow-{\bf n}$
automatically.
The semi-empirical functional form $1/R_c^2\tanh[R_c^2\cdots]$
is applied in Eq.(\ref{Frank2}) to avoid the mathematical 
divergence of the elastic free energy density at the defect centers
and to limit its value to $\Delta f\sim K/R_c^2$, which is the 
correct energy density difference between isotropic and nematic states,
in the defect core regions of size $R_c$.
Another way to avoid the divergence would be to use  
the Landau--de Gennes type free energy with an order parameter 
$
Q_{\alpha\beta}({\bf r})=Q({\bf r})q_{\alpha\beta}({\bf r}),
$
but this requires a prohibitively small lattice spacing near the 
defect points \cite{fukuda}.

The simulation procedure is as follows. i) For a given particle
configuration $\{{\bf R}_1\cdots{\bf R}_N\}$, we obtain
the interface profile $\phi_i({\bf r})$ by Eq.(\ref{profile}). 
Then we can calculate the stable (or meta-stable) nematic configurations
$[q_{\alpha\beta}^{(0)}({\bf r})]$ which satisfy the equilibrium condition
\be
\left.\frac{\delta{\cal F}}{\delta
q_{\alpha\beta}({\bf r})}\right|_{[q_{\alpha\beta}({\bf
 r})]=[q_{\alpha\beta}^{(0)}({\bf r})]}=0
\label{fderiv}
\en
under the director constraint $(n_{\alpha}({\bf r}))^2=1$.
One can perform this by numerical iterations such as the steepest descent or 
the conjugate gradient method.
ii) Once $[q_{\alpha\beta}^{(0)}({\bf r})]$ is obtained, 
the force acting on each particle mediated by the nematic solvents 
follows directly from the Hellmann-Feynman theorem,
\begin{eqnarray}
{\bf f}_i^{PS}&(\{&{\bf R}_1\cdots{\bf R}_N\})\nonumber\\
&=&-\frac{\partial {\cal F}([q_{\alpha\beta}^{(0)}({\bf r})];\{{\bf R}_1\cdots{\bf R}_N\})}{\partial {\bf R}_i}\\
&=&\frac{K}{4R_c^2}\int d{\bf r}\frac{\partial\phi_i}{\partial {\bf R}_i}\tanh\left[R_c^2(\nabla_\alpha q_{\beta \gamma}^{(0)})^2\right]\nonumber\\
&+&W\xi\int d{\bf r}\frac{\partial(\nabla_\alpha\phi_i)}{\partial {\bf R}_i}(\nabla_\beta\phi_i)q_{\alpha\beta}^{(0)} .
\end{eqnarray}
This form is very convenient because we can compute both
$\partial\phi_i/\partial {\bf R}_i$ and 
$\partial(\nabla_\alpha\phi_i)/\partial {\bf R}_i$ at any time
since $\phi_i$ is an analytical function of ${\bf R}_i$.
iii) Finally, we update the particle positions according to
appropriate equations of motion such as 
\be
m_i\frac{d^2{\bf R}_i}{dt^2}
={\bf f}_i^{PP}+{\bf f}_i^{PS}+{\bf f}_i^H+{\bf f}_i^R ,
\label{motion}
\en
where ${\bf f}_i^{PP}$ is the force due to direct particle-particle 
interactions (hard or soft sphere for instance), ${\bf f}_i^{H}$
and ${\bf f}_i^{R}$ are the hydrodynamic and random forces.
Repeating the steps i)$\sim$iii) enables us to perform 
first-principles mesoscopic simulations for the dispersions containing 
many particles without neglecting many-body interactions.

We have performed simple simulations for a two dimensional 
(2D) system to demonstrate our simulation procedure.
The system has $100\times100$ lattice sites in a square box with 
a linear length $L=100$.
Other physical parameters are chosen rather arbitrarily as 
$Rc=1$, $a=5$, and $\xi=2$, where the unit of length is the lattice
spacing $l$.
Since the nematic configurations in 2D can be expressed by a single scalar
field $[\theta({\bf r})]$, the tilt angle of the director against the
horizontal ($x$-) direction,
Eq.(\ref{fderiv}) then reduces to
\be
\frac{\delta{\cal F}}{\delta\theta({\bf r})}
=\frac{\partial q_{\alpha\beta}({\bf r})}
{\partial\theta({\bf r})}\frac{\delta{\cal F}}
{\delta q_{\alpha\beta}({\bf r})}
=0 , 
\en
with $q_{xx}=\cos^2\theta-1/2$, $q_{yy}=\sin^2\theta-1/2$,
and $q_{xy}=q_{yx}=\cos\theta\sin\theta$.
The boundary condition is fixed at $\theta({\bf r})=0$ at the 
edge of the box to avoid rotations of the reference frame.
We first calculated stable nematic configurations around a single 
particle for different $Wa/K$, and found two stable configurations.
The first configuration, which we refer to as weak anchoring, 
contains no topological defect.
In the second configuration, which we refer to as strong anchoring, 
the particle is accompanied by two $-1/2$ charge point defects.
Typical examples of the weak and strong anchoring are shown in Fig.1(a) 
with $Wa/K=2$ and Fig.1(b) with $Wa/K=4$, respectively.
The distance between the defects and the particle center is about $1.3a$.
We note that both configurations possess quadrupolar symmetries, 
and the later would correspond to the Saturn ring configuration 
in three dimensional (3D) systems.
Although in principle particles can be accompanied by one $-1$ 
charge hedgehog defect in 2D as well as in 3D, 
such configurations are unstable in the present 
2D system since the elastic penalty 
of having $m$ point defects with charge $c$ scales as $mKc^2$.
This was directly confirmed by recent simulations with perfect 
normal anchoring \cite{fukuda} and also by our simulations.
The total free energies ${\cal F}/W$ are plotted in Fig.2 as 
functions of $Wa/K$.
While both configurations can coexist in the narrow transition regime 
$2.2\le Wa/K\le2.5$, our model predicts a clear first-order 
transition from the weak anchoring to the strong anchoring around 
$Wa/K\simeq2.3$.

We next simulated the aggregation and ordering process of $30$ colloidal
particles after the isotropic to nematic transition of the solvent occurred.
Here we used the periodic boundary condition and set $Wa/K=4$ so that 
each particle is accompanied by two $-1/2$ charge defects.
Other parameters are the same as in the previous single particle case.
The simulation was performed starting from a random particle
configuration which is a typical configuration when the solvent is in
the isotropic phase ($K=0$). 
We then set $K=1$ and calculated ${\bf f}_i^{PS}$ according to 
the present procedure. 
The particle configurations were updated by 
numerically solving the steepest descent equation,
\be
\zeta\frac{d {\bf R}_i}{dt}={\bf f}_i^{PS}+{\bf f}_i^{PP},
\en
which is obtained by simply substituting 
$d^2{\bf R}_i/dt^2=0$, ${\bf f}_i^R=0$, and 
${\bf f}_i^H=-\zeta d {\bf R}_i/dt$ in Eq.(\ref{motion}).
$\zeta=1$ is a friction constant and thus the off-diagonal components 
of the hydrodynamic interaction was not considered.
Here we obtain ${\bf f}_i^{PP}=-\partial E_{PP}/\partial{\bf R}_i$ 
from the repulsive part of the Lennard-Jones potential, 
$E_{PP}=0.4\sum_{i=1}^{N-1}\sum_{j=i+1}^N\left[\left(2a/|{\bf r}_i-{\bf
r}_j|\right)^{12}-\left(2a/|{\bf r}_i-{\bf r}_j|\right)^{6}+1/4\right]$
truncated at the minimum distance $|{\bf r}_i-{\bf r}_j|=2^{7/6}a$,
to avoid the particles overlapping each other within 
the core radius $\simeq a$.
Snapshots from the present simulation are shown in Fig.3(a) for 
an aggregation stage, and in Fig.3(b) at a later time, where 
the particles are forming ordered clusters due to the 
anisotropic attractions between them.
Note is added that only up to two particle simulations have been done 
so far \cite{stark} and simulations of more than three particles 
would be extremely difficult or almost impossible by other methods 
ever proposed.

In summary, we have developed an extremely powerful simulation method 
to investigate particle dispersions interacting via anisotropic solvents.
We proposed a free energy functional which is suitable for MD type 
simulations. 
The following modifications have been made to the usual Frank free energy
functional.
i) The free energy is given by a functional of a tensor ${\bf q}$ rather 
than a vector ${\bf n}$ to take into account symmetry of the nematic 
director $+{\bf n}\leftrightarrow-{\bf n}$.
ii) The coupling between the nematic solvent and particles at the
interfaces is introduced explicitly through a smooth interface so 
that we can analytically calculate the force acting
on each particle mediated by the host by taking derivatives 
of the free energy according to the particle positions.
iii) The value of the free energy density is limited semi-empirically
to avoid a mathematical divergence in the defect centers.
We have performed demonstrations for a 2D dispersion and confirmed 
that the method works well even when the system contains point defects.
Applications of this method to 3D systems should have no theoretical 
difficulties, but require heavier computation.
This should allow the simulation of the chaining of the particles
caused by the possible dipolar symmetry of the nematic configurations 
around a single particle.
Although we have shown only simple demonstrations of the method by
performing simulations of the 2D system in this letter, simulations 
with physically more interesting situations such as systems with 
non-circular particles, asymmetric particle pairs with different 
particle size, or particles with non-normal anchoring 
as well as more realistic simulations in 3D systems are now underway.

The author thanks Prof. J.P. Hansen and Dr. A. Louis 
for helpful discussions.
He acknowledges also the Ministry of Education, Culture, Sports,
Science and Technology of Japan for supporting his stay in Cambridge 
in 2000/2001.
Calculations have been carried out at the Human Genome Center, 
Institute of Medical Science, University of Tokyo,
and the Supercomputer Center, Institute of Solid State Physics,
University of Tokyo.


\newpage
\narrowtext
\begin{figure}[t]
\centerline{
\epsfxsize=1.7in\epsfbox{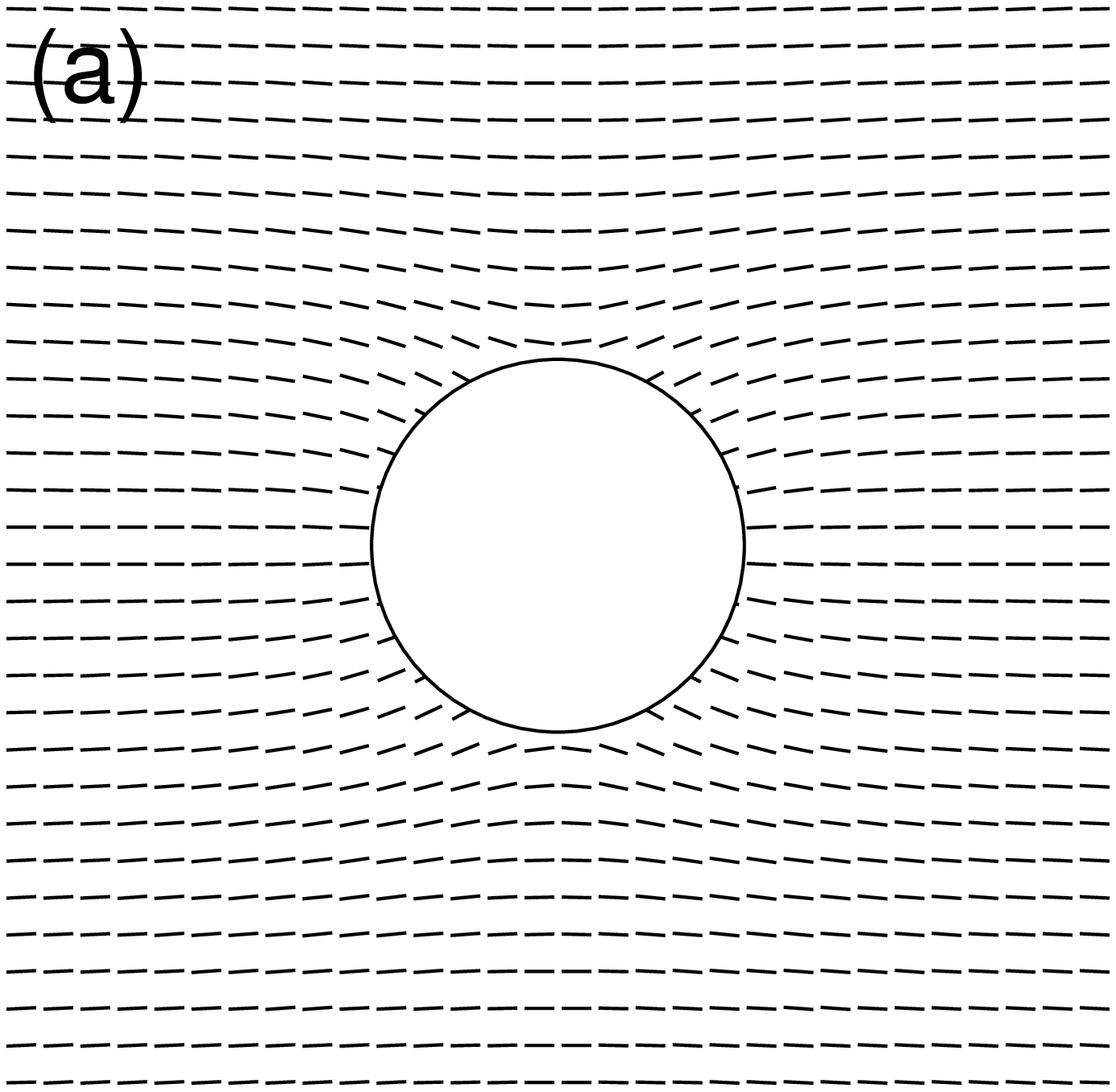}\hspace{1mm}
\epsfxsize=1.7in\epsfbox{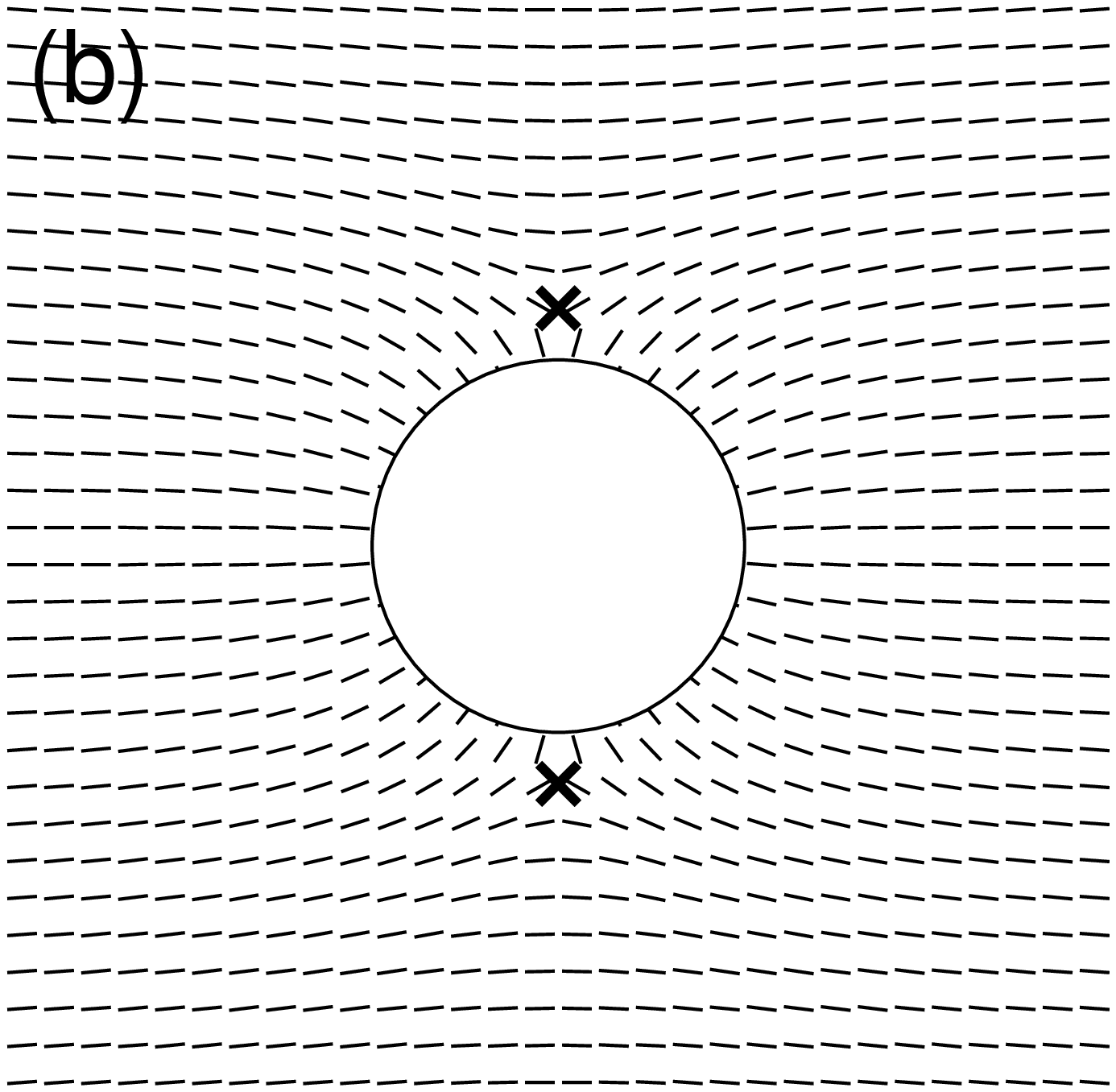}}\vspace{3mm}
\caption{\protect
Director configurations around a single particle for 
(a) weak anchoring case without defect obtained at $Wa/K=2$ and 
(b) strong anchoring case accompanied with two $-1/2$ charge
point defects indicated with the crosses obtained at $Wa/K=4$.
The white disks indicate the particles with radius $a=5$. 
Only $9$\% of the total system is shown for display purpose.
}
\label{fig1}
\end{figure}

\narrowtext
\begin{figure}[b]
\noindent
\centerline{\epsfxsize=3.in\epsfbox{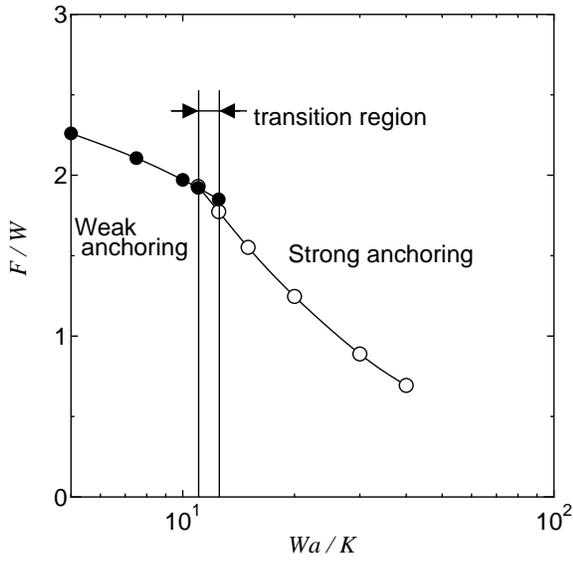}}
\caption{\protect
The total free energies for the single particle cases 
as functions of the strength of the anchoring constant $Wa/K$.
Our model predicts a first-order transition from weak to strong 
anchoring around $Wa/K\simeq2.3$.
}
\label{fig2}
\end{figure}

\narrowtext
\begin{figure}[t]
\centerline{\epsfxsize=3.in\epsfbox{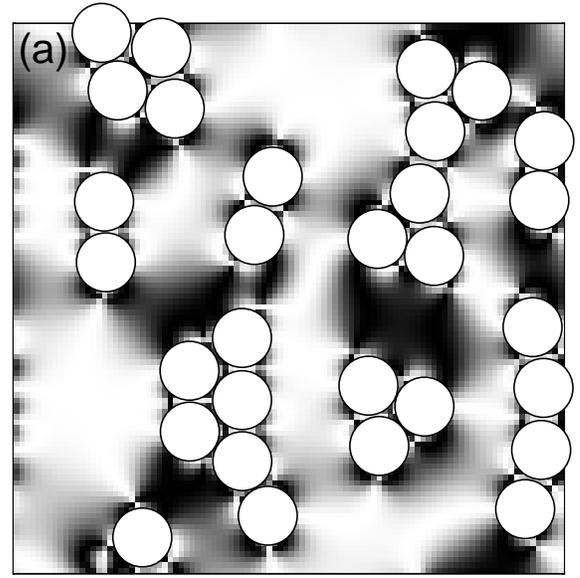}}\vspace{3mm}
\centerline{\epsfxsize=3.in\epsfbox{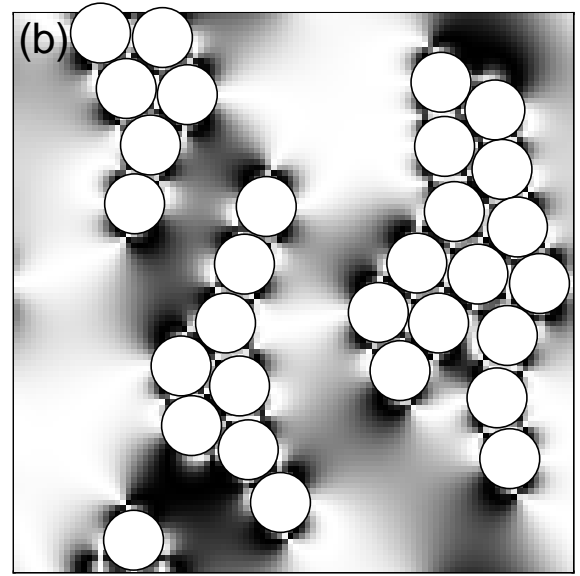}}\vspace{5mm}
\centerline{\epsfxsize=3.in\epsfbox{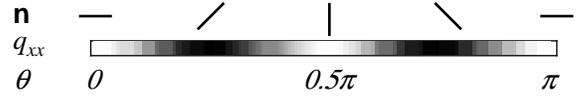}}\vspace{3mm}
\caption{\protect
The aggregation and ordering process of colloidal particles 
when the solvent exhibit the isotropic ($K=0$) to nematic ($K=1$) 
phase transition at $t=0$.
Snapshots (a) in an aggregation stage ($t=10$)
and (b) at a later time ($t=100$), 
where ordering of the particles are observed.
Each particle is accompanied by two $-1/2$ charge point defects.
Darkness presents the value of $q_{xx}^2$.
Black and white correspond to $q_{xx}^2=0$ and $0.25$, respectively.
Those correspond also to $\theta=0.25\pi,0.75\pi$ and $\theta=0,0.5\pi,\pi$ 
as shown in the gradation map.
}
\label{fig3}
\end{figure}

\end{multicols}
\end{document}